# The case for a multi-channel polarization sensitive LIDAR for investigation of insolation-driven ices and atmospheres

Planetary Science Decadal Survey White Paper

Submitted July 10, 2020

**Authors:**
Adrian J. Brown, Plancius Research, MD
adrian.j.brown@nasa.gov
Gorden Videen, Space Science Institute
Evgenij Zubko, Kyung Hee University S. Korea
Nicholas Heavens, Space Science Institute
Nicole-Jeanne Schlegel, Jet Propulsion Lab
Patricio Becerra, University of Bern, Schweiz
Young-Jun Choi, KASI, South Korea
Colin R. Meyer, Dartmouth College
Tanya N. Harrison, Outer Space Institute
Paul Hayne, University of Colorado Boulder
Rachel W. Obbard, SETI Institute
Tim Michaels, SETI Institute
Michael J. Wolff, Space Science Institute
Scott Guzewich, NASA GSFC
Yongxiang Hu, NASA LARC
Claire Newman, Aeolis Research
Christian J. Grund
Chae Kyung Sim, Kyung Hee Uni., S. Korea
Peter B. Buhler, Jet Propulsion Lab
Margaret E. Landis, CU LASP
Timothy J. Stubbs, NASA GSFC
Aymeric Spiga, Sorbonne Uni., LMD, France
Devanshu Jha, MVJCE, India

**Signatories:**
Shane Byrne, LPL University of Arizona
Serina Diniega, Jet Propulsion Lab
Michael Mishchenko, NASA GISS
Sungsoo S. Kim, Kyung Hee Uni., S. Korea
Susan Conway, University of Nantes, France
Ken Herkenhoff, USGS Astrogeology
Michael Mischna, Jet Propulsion Lab
Anthony Colaprete, NASA Ames
Minsup Jeong, KASI, S. Korea
Isaac Smith, York University, Canada and
  Planetary Science Institute, Colorado
Matthew R. Perry, Planetary Science Institute
John E. Moores, York University, Canada
Christine S. Hvidberg, Niels Bohr Inst., Denmark
Jonathan A. R. Rall, Off. of Chf Sci., NASA HQ
Sylvain Piqueux, Jet Propulsion Lab
Robert Lillis, SSL UC Berkeley
Leslie Tamppari, Jet Propulsion Lab
Wendy Calvin, University of Nevada
Jennifer Hanley, Lowell Observatory
Nathaniel Putzig, Planetary Science Institute
Ali M. Bramson, Purdue University
Bonnie Meineke, Ball Aerospace
Michael Veto, Ball Aerospace
Jack Holt, University of Arizona
Bryana L. Henderson, Jet Propulsion Lab
Lori K. Fenton, SETI Institute
Alain Khayat, NASA GSFC/UMD
Tim McConnochie, UMD/Space Science Inst.
Timothy N. Titus, USGS Astrogeology

> **"There once was a LIDAR with Polarized Holes,**
> **Made to observe Mars' Mysterious Poles,**
> **It measured Volatiles and Ices ...**
> **and used Multispectral Devices ...**
> **And met many of our Mars Climate Goals." - Anon. 2020**

**Key point of this white paper:** All LIDAR instruments are not the same, and advancement of LIDAR technology requires an ongoing interest and demand from the community to foster further development of the required components. The purpose of this paper is to make the community aware of the need for further technical development, and the potential payoff of investing experimental time, money and thought into the next generation of LIDARs.

**Technologies for development:** We advocate for future development of LIDAR technologies to measure the **polarization** state of the reflected light at **selected multiple wavelengths,** chosen according to the species of interest (e.g., $H_2O$ and $CO_2$ in the Martian setting).

**Key scientific questions:** In the coming decade, dollars spent on these LIDAR technologies will go towards addressing key climate questions on Mars and other rocky bodies, particularly those with seasonally changing (i.e. insolation driven) plumes of multiple icy volatiles such as Mars, Enceladus, Triton, or Pluto, and insolation-driven dust lifting, such as cometary bodies and the Moon. We will show from examining past Martian and terrestrial lidars that orbital and landed LIDARs can be effective for producing new insights into insolation-driven processes in current planetary climate on several bodies, beyond that available to our current fleet of largely passive instruments on planetary missions.

## 1. Preamble and Science Themes

This white paper is intended to make the decadal survey panel aware of the type and amount of information about the present day Martian climate that would be acquired by a multiple-wavelength, active, near-infrared (NIR) LIDAR instrument. This instrument would measure the reflected intensity and polarization characteristics of backscattered radiation from the surface via the atmosphere, such as that described by Brown et al. [1]. Science output from this type of dataset would address the following three major science themes:

**Science Theme 1. Surface** Global, night and day mapping of $H_2O$ and $CO_2$ surface ice,

**Science Theme 2. Ice Clouds**: Unambiguous discrimination and seasonal mapping of $CO_2$ and $H_2O$ ice clouds

**Science Theme 3. Dust Aerosols**: Inference of dust grain shapes and size distributions from multiwavelength polarization measurements.

Knowledge generated from such an instrument has the potential to fundamentally shift our understanding of modern-day Martian volatile transport and deposition. As a bonus, this lidar would permit a continuation of the MOLA geodetic mapping of Mars to a point where a global network suitable for registration of high resolution images (e.g. HiRISE).

## 2. Martian Example Case Study LIDAR - *ASPEN*

Our present understanding of the sublimation of surface $H_2O$ and $CO_2$ ices and related atmospheric changes on Mars is the result of recent polewide and seasonal studies of springtime recession using the Mars Climate Sounder [2], CRISM [3], and MARCI [4] instruments on MRO, the OMEGA instrument on Mars Express [5-6], the THEMIS instrument on 2001 Mars Odyssey [7] and the TES instrument on Mars Global Surveyor [8]. These investigations have steadily advanced our understanding of major polar processes. However, the observations of the spatially localized springtime recession phenomena, such as geysers (gas/dust jets, which are inferred, and have not been directly observed) [9], and observations of the asymmetric retraction of the seasonal cap [6,10] lead us to ask a key scientific question – what role does spatially localized and temporally intermittent deposition of ices and dust during fall and winter [10] play in the annual $CO_2$ and $H_2O$ cycles that are instrumental the climate of modern-day Mars? In the remainder of this white paper, we discuss an instrument optimized for Martian conditions called "Atmospheric/Surface Polarization Experiment at Nighttime" (ASPEN) [1], designed to directly address the role of ice and dust deposition in Martian climate.

*Instrument Specs*. ASPEN would be a multi-wavelength, altitude-resolved, active, near-infrared (NIR: covering 1.52-1.59 μm) instrument to measure the reflected intensity and polarization of backscattered radiation from planetary surfaces and atmospheres. The currently envisioned spacecraft instrument utilizes multiple diode lasers, each operable at a different wavelength, amplified by a fiber laser stage (commonly referred to as a master-oscillator power-amplifier or MOPA). The laser would operate in a high repetition (10 kHz) low pulse energy (40μJ) configuration. The receiver side will consist of a telescope coupled to an indium gallium arsenide (InGaAs) high dynamic range avalanche photo detector (APD).

*Martian Operations*. As currently envisioned, the ASPEN instrument would operate as a line profile instrument, in a similar manner to the MOLA lidar. The instrument is best suited for an MGS or MRO-type 250–320 km elliptical/circular orbit but could also operate in an elliptical orbit with reduced sensitivity during apoapsis. For optimized polar measurements, orbital inclination should be between 85° and 92.8°. An elliptical orbit such as that mentioned in the recent MSO SAG document would allow lidar-occultation measurements of the atmosphere, allowing the atmosphere to be viewed 'side on', thus enabling profile measurements of $CO_2$, $H_2O$ ice and vapor in the Martian atmosphere.

Preliminary instrument performance calculations of common measurement scenarios for the MOPA laser ASPEN instrument estimate the surface spot size at ~25m on the surface and a horizontal resolution of ~275m (similar to MOLA which was ~330m).

*Multi-wavelength*. In order to take advantage of the tremendous research and development that has gone into lasers and fiber optic components that operate in the near-IR *by the telecommunications industry* in recent years, the instrument will operate at wavelengths between 1.52 and 1.59 μm. These wavelengths are ideally suited to discriminate $CO_2$ and $H_2O$ ices and vapor using the differential absorption lidar (DIAL) technique originally developed for terrestrial remote sensing [26, 27].

The Mars Science community has previously recognized the need for an ASPEN-type instrument. The need for actively scanning laser sensors that operate over a range of frequencies was acknowledged in the 2006 Solar System Exploration Roadmap [11] (page 108). In addition, the 2013 Mars Science Orbiter Science Analysis Group (MSO SAG) report stated that a "multibeam lidar" similar to the LOLA on LRO, and inheriting many aspects from the CALIPSO LIDAR, would "resolve optically dense atmospheric phenomena" and "significantly constrain seasonal mass budgets". In essence, it was thought to be an ideal instrument for a "2013 MSO mission" [12].

*Key Science:* Previous instruments have given glimpses of cloud and surface ice activity on Mars, but no previous Martian orbital instrument has been able to simultaneously address the following science measurements, which ASPEN would:

a.) Detect clouds up to 100km above the Martian surface during night and day;

b.) Discriminate between $H_2O$ and $CO_2$ ice and dust on the surface and in aerosols in the atmosphere in both polar day and night [13];

c.) Map cloud structure using lidar backscatter and depolarization;

d.) Map large-grained (path length up to 30cm) $CO_2$ slab ice in the polar night, which is uniquely Martian and is extremely poorly understood [14];

e.) Determine whether the $H_2O$ ice signature in the southern polar trough system is due to cloud [15] or surface ice [16];

f.) Monitor 'cold spot' activity during the polar night and determine whether these enigmatic features are due to $CO_2$ clouds, precipitation or surface ice [17,18];

g.) Monitor night and day gas/dust jet (geyser) activity within the 'Cryptic Region' (which have not yet been observed "in action") in southern late winter and early spring and determine what amount of solar energy is required for them to be active [8,19];

h.) Uniquely identify cloud types and platelet/grain orientation, in order to confirm the presence and structure of convective $CO_2$ cloud towers, a potentially critical part of the polar night dynamics and energy partitioning [20];

i.) Atmospheric column dust optical depths when instrument is in operation [21,22].

j.) Monitor the spring and summertime retreating polar caps for signs of sediment flows and possibly even geysers caused by subliming $CO_2$ ice. This type of activity has already been suggested to cause changes on mid-latitude gullies [23] and dunes [24].

k.) Address questions of spatial extent (locality and 'deep transport') of Martian cloud structure, which is anticipated to be on the order of 1km width and is crucial to understanding differences between terrestrial and Martian mesospheric dynamics [25].

i.) conduct a sensitive global (daytime and nighttime) search for outbursts of Martian methane down to 2ppb with relative errors of 0.5% [26,27].

## 3. Instrument Concept and Background

### *3.1 A conceptual model for energy transport in the Martian climate and why a lidar is essential to reveal it*

Our understanding of the modern day Martian climate is based on relatively easy to visualize energy transport concepts. One can envisage the energy within the polar caps as governed by the latent heat of the ice in the caps, and can track the flow of ice from the caps to the atmosphere and mid latitudes during summer. These processes can be observed as the springtime recession of the seasonal cap, as the stored energy of the ice is broken up and moves into the atmosphere and the mid latitudes. Energy transport in the Martian climate has been tracked by telescopes and subsequent orbital instruments.

During winter, the reverse process is less certain, mainly because it can no longer be directly observed due to the onset of the polar hood and polar night [3]. It is during this time that the energy of the cap, like ice blocks in an igloo, is stored away again in the form of $CO_2$ and $H_2O$ ice in the winter cap. However, this process is not well understood at all - we have some clues in the southern pole due to supercool spots that have been associated with $CO_2$ precipitation, and we know that during this time large blocks of $CO_2$ ice also form in order to create the araneiform terrain visible during springtime. However, their formation processes remain a mystery due to the inability of our passive imaging and spectroscopic instruments to pierce the cloudy enveloping veil. Further, it should be noted that radar instruments are generally not sensitive to the top 1mm-10cm of the ice, where the daily deposition and sublimation processes occur.

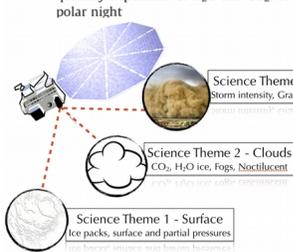

Figure 1. ASPEN instrument science themes, measurement objective and instrument requirements [1].

*3.2. Previous Lidar Mission – MOLA*

The highly successful Mars Orbiting Laser Altimeter (MOLA) instrument on Mars Global Surveyor measured clouds and the height of the seasonal $CO_2$ surface ice accumulations [28,29]. However, its use of a single wavelength (1.064µm) prevented discrimination between $H_2O$ and $CO_2$ clouds using the MOLA dataset. In addition, MOLA had no ability to assess particle sizes/shapes, nor measure $H_2O$ or $CO_2$ gas vapor.

The MOLA instrument did demonstrate the ability to detect optically thin Martian dust devils [30,31]. Consequently, one can have confidence that ASPEN will be capable of monitoring dust loading and activity, including that associated with the eruption of 'geysers' in the south polar 'Cryptic' Region [8] - because the ASPEN detectors are designed not to saturate over the relatively high albedo Martian ice caps [32].

*3.3. Previous Lidar Mission – Phoenix*

The Phoenix spacecraft landed in the Vastitas Borealis region of Mars (at 68.2° N) in May 2008 and operated for 5 months or 152 Martian days (one summer and fall period) [33]. The Phoenix meteorology station included a vertical pointing Nd:YAG lidar operating at 1.064 and 0.532 µm. The lidar system successfully detected aerosol structures consistent with Martian cirrus clouds and in particular the 'virga' or "Mare's Tails" (ice particles falling from their formation site in the main cloud deck) as they passed over the lander during the local night [34]. Having no polarization capability, the Phoenix lidar could not directly determine grain shapes. We consider the Phoenix lidar to be a useful pathfinder for future lidar systems such as ASPEN.

*3.4 Previous Terrestrial LIDAR mission - CALIPSO*

The CALIOP laser onboard the Cloud Aerosol Lidar and Infrared Pathfinder Satellite Observations (CALIPSO) space-craft was launched in April 2006 and is still in operation. With an orbit ~700 km, it is part of the 'A-Train' of Earth observing satellites. The CALIOP laser operates at 1.064 and 0.532 µm, measuring linear polarization in the latter band. The instrument was designed and tested at Ball Aerospace and is operated jointly by NASA and CNES [35].

The surface footprint of the CALIOP is ~100 m and the vertical resolution is 30–60 m. An example application of the CALIPSO instrument was observation of soot from Arctic wildfires drifting over Greenland [36]. The sensitivity to the aerosols associated with the fires provides a clear demonstration of lidar utility for monitoring/characterizing dust and cloud activity across multiple scales, as well as for studies of low lying fogs and sublimation flow events near the Martian surface. CALIPSO has also been used to monitor the height of the terrestrial planetary boundary layer, a useful precursor experiment for ASPEN at Mars. Although CALIOP does not exhibit the same wavelength flexibility and polarimetric capability (i.e. does not measure the full returned Stokes vector) of the ASPEN instrument, its enhanced abilities beyond the MOLA and Phoenix lidar provide further motivation for the concept of an orbital lidar at Mars.

## 4. Current TRL and Future technology advancement

The ASPEN concept is currently at TRL 2-3, with the initial designs and requirements laid out [1] and work on multiwavelength lasers [26]. Work on enabling technologies for variable frequency laser sources and detectors at GSFC has demonstrated the ability of measurement of $CH_4$ in the Earth's atmosphere using a wavelength manifold (rather than individual rotational lines) which has enabled accuracies of 0.5% [26]. In addition, work continues on a Martian wind-measuring LIDAR, called MARLI [37], which is complementary to the ASPEN concept. The MARLI design is a single wavelength approach that concentrates on atmospheric winds and dust and water ice clouds, in contrast to our multiwavelength, surface and atmosphere approach.

We anticipate a 3 year, $1 million project would get the instrument to TRL 4 with a breadboard done and retire much of the risk around the laser and amplifier. To advance the instrument from TRL 4 to TRL 6, with an instrument in a terrestrial airborne environment, would take roughly $3-5 million and a further 3-4 years of development. At that stage, the instrument could be proposed for future spaceflight missions.

## 5. Application to other planetary bodies

In this white paper, we have emphasized the utility of the ASPEN lidar instrument for an orbital Mars mission; however the same type of instrument would be applicable for a range of future missions. The multiwavelength polarization sensitive LIDAR would also provide invaluable insights when deployed to other rocky bodies, particularly those with seasonally changing (i.e. insolation driven) plumes of multiple icy volatiles, such as Enceladus, Triton, or Pluto, and insolation driven dust lifting, such as Ceres, Main Belt Comets and the Moon.

The ASPEN instrument concept would be ideal for missions to ice covered bodies (e.g. Europa, Enceladus, Triton, even methane ice on Kuiper Belt objects) to investigate the properties of icy surfaces in low sunlight conditions. As part of a Discovery class mission to active comet-like near Earth asteroids, the system would be ideal for probing the physical properties of a cometary coma. The instrument could also be used in an orbital mission to Venus,to probe cloud properties and structure in the NIR windows of the Venusian atmosphere.

## 6. Summary

We have outlined the scientific case for a polarization LIDAR for an eventual orbital mission to Mars. The combination of active, multiple-wavelength measurements with polarimetry makes this instrument concept an essential option in the future inventory of spacecraft instrumentation. The lessons learned from such an instrument would fundamentally advance our understanding of modern day volatile transport and deposition on Mars, and other planetary surfaces with insolation-driven volatile regimes.

**References**

[1] Brown, Adrian J., Timothy I. Michaels, Shane Byrne, Wenbo Sun, Timothy N. Titus, Anthony Colaprete, Michael J. Wolff, and Gorden Videen. "The Science Case for a Modern, Multi-Wavelength, Polarization-Sensitive LIDAR in Orbit around Mars." *Journal of Quantitative Spectroscopy & Radiative Transfer*, no. 153 (2015): 131–43. https://doi.org/10.1016/j.jqsrt.2014.10.021.

[2] Hayne, Paul O., David A. Paige, John T. Schofield, David M. Kass, Armin Kleinbˆhl, Nicholas G. Heavens, and Daniel J. McCleese. "Carbon Dioxide Snow Clouds on Mars: South Polar Winter Observations by the Mars Climate Sounder." *Journal of Geophysical Research* 117, no. E8 (2012): E08014. http://dx.doi.org/10.1029/2011JE004040.

[3] Brown, Adrian J., Shane Byrne, Livio L. Tornabene, and Ted Roush. "Louth Crater: Evolution of a Layered Water Ice Mound." *Icarus* 196, no. 2 (2008): 433–45. http://arxiv.org/abs/1401.8024.

Brown, Adrian J., Wendy M. Calvin, Patrick C. McGuire, and Scott L. Murchie. "Compact Reconnaissance Imaging Spectrometer for Mars (CRISM) South Polar Mapping: First Mars Year of Observations." *Journal of Geophysical Research* 115 (2010): doi:10.1029/2009JE003333. http://arxiv.org/abs/1402.0541.

Brown, Adrian J., Wendy M. Calvin, and Scott L. Murchie. "Compact Reconnaissance Imaging Spectrometer for Mars (CRISM) North Polar Springtime Recession Mapping: First 3 Mars Years of Observations." *Journal of Geophysical Research* 117 (2012): E00J20. http://dx.doi.org/10.1029/2012JE004113.

Brown, Adrian J., Sylvain Piqueux, and Timothy N. Titus. "Interannual Observations and Quantification of Summertime H2O Ice Deposition on the Martian CO2 Ice South Polar Cap." *Earth and Planetary Science Letters* 406 (November 15, 2014): 102–9. https://doi.org/10.1016/j.epsl.2014.08.039.

Bapst, Jonathan, Shane Byrne, and Adrian J. Brown. "On the Icy Edge at Louth and Korolev Craters." *Icarus*, Mars Polar Science VI, 308 (July 1, 2018): 15–26. https://doi.org/10.1016/j.icarus.2017.10.004.

[4] Wolff, Michael J., R. Todd Clancy, Jay D. Goguen, Michael C. Malin, and Bruce A. Cantor. "Ultraviolet Dust Aerosol Properties as Observed by MARCI." *Icarus* 208, no. 1 (2010): 143–55. http://www.sciencedirect.com/science/article/pii/S0019103510000205.

[5] Langevin, Y., S. Douté, M. Vincendon, F. Poulet, J.-P. Bibring, B. Gondet, B. Schmitt, and F. Forget. "No Signature of Clear CO2 Ice from the 'cryptic' Regions in Mars' South Seasonal Polar Cap." *Nature* 442, no. 7104 (2006): 790–92.

[6] Appere, T., B. Schmitt, Y. Langevin, S. DoutÈ, A. Pommerol, F. Forget, A. Spiga, B. Gondet, and J. P. Bibring. "Winter and Spring Evolution of Northern Seasonal Deposits on Mars from OMEGA on Mars Express." *Journal of Geophysical Research* 116, no. E5 (2011): E05001. http://dx.doi.org/10.1029/2010JE003762.

[7] Titus, T. N., H. H. Kieffer, and P. R. Christensen. "Exposed Water Ice Discovered near the South Pole of Mars." *Science* 299, no. 5609 (February 14, 2003): 1048–51. ://000180960000042.

[8] Kieffer, H.H., and T.N. Titus. "TES Mapping of Mars' North Seasonal Cap." *Icarus* 154, no. 1 (2001): 162–80.

[9] Kieffer, H.H., P.R. Christensen, and T.N. Titus. "CO2 Jets Formed by Sublimation beneath Translucent Slab Ice in Mars' Seasonal South Polar Ice Cap." *Nature* 442 (2006): 793–96.

[10] Brown, Adrian J., Wendy M. Calvin, Patricio Becerra, and Shane Byrne. "Martian North Polar Cap Summer Water Cycle." *Icarus* 277 (2016): 401–15. https://doi.org/10.1016/j.icarus.2016.05.007.

[11] NASA. "Solar System Exploration – 2006 Solar System Exploration Roadmap." NASA, 2006.

[12] Calvin, W., M. Allen, W. B. Banerdt, D. Banfield, B.A. Campbell, P.R. Christensen, K.S. Edgett, et al. "Report from the 2013 Mars Science Orbiter (MSO) Second Science Analysis Group." Pasadena, CA: JPL, May 29, 2007.

[13] Gary–Bicas, C. E., P. O. Hayne, T. Horvath, N. G. Heavens, D. M. Kass, A. Kleinböhl, S. Piqueux, J. H. Shirley, J. T. Schofield, and D. J. McCleese. "Asymmetries in Snowfall, Emissivity, and Albedo of Mars' Seasonal Polar Caps: Mars Climate Sounder Observations." *Journal of Geophysical Research: Planets* 125, no. 5 (2020): e2019JE006150. https://doi.org/10.1029/2019JE006150.

Becerra, Patricio, Shane Byrne, and Adrian J. Brown. "Transient Bright 'Halos' on the South Polar Residual Cap of Mars: Implications for Mass-Balance." *Icarus*, no. 251 (2015): 211–25. https://doi.org/10.1016/j.icarus.2014.04.050.

[14] Langevin, Y., J.-P. Bibring, F. Montmessin, F. Forget, M. Vincendon, S. Douté, F. Poulet, and B. Gondet. "Observations of the South Seasonal Cap of Mars during Recession in 2004–2006 by the OMEGA Visible/near-Infrared Imaging Spectrometer on Board Mars Express." *Journal of Geophysical Research* 112, no. E08S12 (2007): 10.1029/2006JE002841. 10.1029/2006JE002841.

[15] Inada, Ai, Mark I. Richardson, Timothy H. McConnochie, Melissa J. Strausberg, Huiqun Wang, and James F. Bell Iii. "High-Resolution Atmospheric Observations by the Mars Odyssey Thermal Emission Imaging System." *Icarus* 192, no. 2 (2007): 378–95. http://www.sciencedirect.com/science/article/B6WGF-4PMT31G-1/2/4df14f5addeac3580cf3f98e587d1d48.

[16] Titus, T.N. "Thermal Infrared and Visual Observations of a Water Ice Lag in the Mars Southern Summer." *Geophysical Research Letters* 32, no. L24204 (2005): doi:10.1029/2005GL024211. doi:10.1029/2005GL024211.

[17] Forget, F., and J. B. Pollack. "Thermal Infrared Observations of the Condensing Martian Polar Caps: CO2 Ice Temperatures and Radiative Budget." *Journal of Geophysical Research-Planets* 101, no. E7 (July 25, 1996): 16865–79. ://A1996UZ72800004.

[18] Hayne, Paul O., David A. Paige, and Nicholas G. Heavens. "The Role of Snowfall in Forming the Seasonal Ice Caps of Mars: Models and Constraints from the Mars Climate Sounder." *Icarus* 231, no. 0 (2013): 122–30. http://www.sciencedirect.com/science/article/pii/S0019103513004491.

[19] Piqueux, Sylvain, and Philip R. Christensen. "Deposition of CO2 and Erosion of the Martian South Perennial Cap between 1972 and 2004: Implications for Current Climate Change." *Journal of Geophysical Research* 113, no. E02006 (2008): doi:10.1029/2007JE002969. doi:10.1029/2007JE002969.

[20] Colaprete, Anthony, Robert M. Haberle, and Owen B. Toon. "Formation of Convective Carbon Dioxide Clouds near the South Pole of Mars." *Journal of Geophysical Research* 108 (2003): doi://10.1029/2003JE002053. http://dx.doi.org/10.1029/2003JE002053.

[21] Ma, Yingying, Wei Gong, Pucai Wang, and Xiuqing Hu. "New Dust Aerosol Identification Method for Spaceborne Lidar Measurements." *Journal of Quantitative Spectroscopy and Radiative Transfer* 112, no. 2 (2011): 338–45. http://www.sciencedirect.com/science/article/B6TVR-50S8PW0-1/2/334ef6faa4230a5c293bf7b14eaecee4.

[22] Lu, Xiaomei, Yuesong Jiang, Xuguo Zhang, Xuan Wang, and Nicola Spinelli. "Two-Wavelength Lidar Inversion Algorithm for Determination of Aerosol Extinction-to-Backscatter Ratio and Its Application to CALIPSO Lidar Measurements." *Journal of Quantitative Spectroscopy and Radiative Transfer* 112, no. 2 (2010): 320–28. http://www.sciencedirect.com/science/article/B6TVR-50PCMKF-1/2/56222207ee5ef323d7f0dd4fe55163f7.

[23] Raack, Jan, Susan J. Conway, Thomas Heyer, Valentin T. Bickel, Meven Philippe, Harald Hiesinger, Andreas Johnsson, and Marion Massé. "Present-Day Gully Activity in Sisyphi Cavi, Mars – Flow-like Features and Block Movements." *Icarus* 350 (November 1, 2020): 113899. https://doi.org/10.1016/j.icarus.2020.113899.

[24] Diniega, Serina, Shane Byrne, Nathan T. Bridges, Colin M. Dundas, and Alfred S. McEwen. "Seasonality of Present-Day Martian Dune-Gully Activity." *Geology* 38, no. 11 (November 1, 2010): 1047–50. https://doi.org/10.1130/G31287.1.

Diniega, Serina, Candice J. Hansen, Amanda Allen, Nathan Grigsby, Zheyu Li, Tyler Perez, and Matthew Chojnacki. "Dune-Slope Activity Due to Frost and Wind throughout the North Polar Erg, Mars." *Geological Society, London, Special Publications* 467, no. 1 (January 1, 2019): 95–114. https://doi.org/10.1144/SP467.6.

[25] Michaels, T. I., A. Colaprete, and S. C. R. Rafkin. "Significant Vertical Water Transport by Mountain-Induced Circulations on Mars." *Geophysical Research Letters* 33, no. 16 (August 1, 2006): L16201. https://doi.org/10.1029/2006GL026562.

Rafkin, Scot C. R. "The Potential Importance of Non-Local, Deep Transport on the Energetics, Momentum, Chemistry, and Aerosol Distributions in the Atmospheres of Earth, Mars and Titan." *Planetary and Space Science* 60, no. 1 (2011): 147–54. http://www.sciencedirect.com/science/article/pii/S0032063311002364.

Heavens, Nicholas G., David M. Kass, James H. Shirley, Sylvain Piqueux, and Bruce A. Cantor. "An Observational Overview of Dusty Deep Convection in Martian Dust Storms." *Journal of the Atmospheric Sciences* 76, no. 11 (November 1, 2019): 3299–3326. https://doi.org/10.1175/JAS-D-19-0042.1.

[26] Riris, Haris, Kenji Numata, Stewart Wu, Jes Sherman, Gordon Morrison, Henry Garrett, and Milan L. Mashanovitch. "A New Laser Transmitter for Methane and Water Vapor Measurements at 1.65 Mm." In *Laser Radar Technology and Applications XXV*, 11410:1141007. International Society for Optics and Photonics, 2020. https://doi.org/10.1117/12.2558816.

[27] Riris, Haris, Kenji Numata, Stewart Wu, and Molly Fahey. "The Challenges of Measuring Methane from Space with a LIDAR." *CEAS Space Journal* 11, no. 4 (December 1, 2019): 475–83. https://doi.org/10.1007/s12567-019-00274-8.

[28] Smith, D. E., M. T. Zuber, S. C. Solomon, R. J. Phillips, J. W. Head, J. B. Garvin, W. B. Banerdt, et al. "The Global Topography of Mars and Implications for Surface Evolution." *Science* 284, no. 5419 (May 28, 1999): 1495–1503. ://000080548100029.

[29] Pettengill, Gordon H., and Peter G. Ford. "Winter Clouds Over the North Martian Polar Cap." *Geophysical Research Letters* 27, no. 5 (2000): 609–612.

[30] Ivanov, Anton B., and Duane O. Muhleman. "Cloud Reflection Observations: Results from the Mars Orbiter Laser Altimeter." *Icarus* 154, no. 1 (2001): 190–206. http://www.sciencedirect.com/science/article/B6WGF-457CXN8-F/2/361d6664c658462dc2a32c8ace097028.

[31] Neumann, Gregory A., David E. Smith, and Maria T. Zuber. "Two Mars Years of Clouds Detected by the Mars Orbiter Laser Altimeter." *Journal of Geophysical Research* 108, no. E4 (2003). doi:10.1029/2002JE001849.

[32] Heavens, N. G. "The Reflectivity of Mars at 1064 Nm: Derivation from Mars Orbiter Laser Altimeter Data and Application to Climatology and Meteorology." *Icarus* 289 (June 2017): 1–21. https://doi.org/10.1016/j.icarus.2017.01.032.

[33] Smith, P. H., L. K. Tamppari, R. E. Arvidson, D. Bass, D. Blaney, W. V. Boynton, A. Carswell, et al. "H2O at the Phoenix Landing Site." *Science* 325, no. 5936 (2009): 58–61. http://www.sciencemag.org/cgi/content/abstract/325/5936/58.

[34] Whiteway, J. A., L. Komguem, C. Dickinson, C. Cook, M. Illnicki, J. Seabrook, V. Popovici, et al. "Mars Water-Ice Clouds and Precipitation." *Science* 325, no. 5936 (July 3, 2009): 68–70. https://doi.org/10.1126/science.1172344.

[35] Weimer, C. A., L. Ruppert, and J. Spelman. "Commissioning of the CALIPSO Payload," Vol. 6555. San Diego, CA: SPIE, 2007.

[36] Box, Jason, Marco Tedesco, Xavier Fettweis, Dorothy Hall, Konrad Steffen, and Julienne Stroeve. "Greenland Ice Sheet Albedo Feedback: Mass Balance Implications." San Francisco, CA, 2012.

[37] Cremons, Daniel R., James B. Abshire, Xiaoli Sun, Graham Allan, Haris Riris,

Michael D. Smith, Scott Guzewich, Anthony Yu, and Floyd Hovis. "Design of a Direct-Detection Wind and Aerosol Lidar for Mars Orbit." *CEAS Space Journal* 12, no. 2 (June 2020): 149–62. https://doi.org/10.1007/s12567-020-00301-z.